\documentclass[10pt,twocolumn]{article}
\usepackage[utf8]{inputenc}
\usepackage{extsizes}
\usepackage[super,sort&compress,comma]{natbib} 
\usepackage[version=3]{mhchem}
\usepackage[left=1.5cm, right=1.5cm, top=1.785cm, bottom=2.0cm]{geometry}
\usepackage{balance}
\usepackage{mathptmx}
\usepackage{sectsty}
\usepackage{amsmath}
\usepackage{bm}
\usepackage{graphicx} 
\usepackage{lastpage}
\usepackage[format=plain,justification=justified,singlelinecheck=false,font={stretch=1.125,small,sf},labelfont=bf,labelsep=space]{caption}
\usepackage{float}
\usepackage{fancyhdr}
\usepackage{fnpos}
\usepackage[english]{babel}
\addto{\captionsenglish}{%
  
}
\usepackage{array}
\usepackage{droidsans}
\usepackage{charter}
\usepackage[T1]{fontenc}
\usepackage[usenames,dvipsnames]{xcolor}
\usepackage{setspace}
\usepackage{siunitx}
\usepackage[compact]{titlesec}
\usepackage[hidelinks]{hyperref}
\usepackage{epstopdf}%This line makes .eps figures into .pdf - please comment out if not required.

\definecolor{cream}{RGB}{222,217,201}
\begin{document}
\title{Self-Driving Laboratory Optimizes the Lower Critical Solution Temperature of Thermoresponsive Polymers}
\author{Guoyue Xu \and Renzheng Zhang \and Tengfei Luo\thanks{Corresponding author: tluo@nd.edu}\\
\small Department of Aerospace and Mechanical Engineering, University of Notre Dame, IN 46556, USA}

\date{\vspace{-\baselineskip}}
\maketitle
\makeatletter 
\newlength{\figrulesep} 
\setlength{\figrulesep}{0.5\textfloatsep} 

\newcommand{\topfigrule}{\vspace*{-1pt}% 
\noindent{\color{cream}\rule[-\figrulesep]{\columnwidth}{1.5pt}} }

\newcommand{\botfigrule}{\vspace*{-2pt}% 
\noindent{\color{cream}\rule[\figrulesep]{\columnwidth}{1.5pt}} }

\newcommand{\dblfigrule}{\vspace*{-1pt}% 
\noindent{\color{cream}\rule[-\figrulesep]{\textwidth}{1.5pt}} }

\makeatother
%%%END OF FIGURE SETUP%%%

%%%TITLE AND AUTHORS%%%
% \twocolumn[
%   \begin{@twocolumnfalse}
% {\includegraphics[height=30pt]{head_foot/journal_name}\hfill\raisebox{0pt}[0pt][0pt]{\includegraphics[height=55pt]{head_foot/RSC_LOGO_CMYK}}\\[1ex]
% \includegraphics[width=18.5cm]{head_foot/header_bar}}\par
% \vspace{1em}
% \sffamily
% \begin{tabular}{m{4.5cm} p{13.5cm} }

% \includegraphics{head_foot/DOI} & \noindent\LARGE{\textbf{Self-Driving Laboratory Optimizes the Lower Critical Solution Temperature of Thermoresponsive Polymers$^\dag$}} \\%Article title goes here instead of the text "This is the title"
%  & \vspace{0.3cm} \\

%  & \noindent\large{Guoyue Xu,$^{\ast}$\textit{$^{a}$} Renzheng Zhang,\textit{$^{a}$} and Tengfei Luo\textit{$^{a}$}} \\%Author names go here instead of "Full name", etc.

% \includegraphics{head_foot/dates} & \\

% \end{tabular}

%  \end{@twocolumnfalse} \vspace{0.6cm}

%   ]
%%%END OF TITLE AND AUTHORS%%%

%%%FONT SETUP - please do not change any commands within this section
\renewcommand*\rmdefault{bch}\normalfont\upshape
\rmfamily
\section*{}
\vspace{-1cm}

%%%FOOTNOTES%%%

% \footnotetext{\textit{$^{a}$~ Department of Aerospace and Mechanical Engineering, University of Notre Dame, IN 46556 USA.}}

%Please use \dag to cite the ESI in the main text of the article.
%If you article does not have ESI please remove the the \dag symbol from the title and the footnotetext below.
% \footnotetext{\dag~Supplementary Information available: [details of any supplementary information available should be included here]. See DOI: 00.0000/00000000.}
%additional addresses can be cited as above using the lower-case letters, c, d, e... If all authors are from the same address, no letter is required

%%%END OF FOOTNOTES%%%

%%%ABSTRACT%%%%

\sffamily{\textbf{To overcome the inherent inefficiencies of traditional trial-and-error materials discovery, the scientific community is increasingly developing autonomous laboratories that integrate data-driven decision-making into closed-loop experimental workflows. In this work, we realize this concept for thermoresponsive polymers by developing a low-cost, "frugal twin" platform for the optimization of the lower critical solution temperature (LCST) of poly(N-isopropylacrylamide) (PNIPAM). Our system integrates robotic fluid-handling, on-line sensors, and Bayesian optimization (BO) that navigates the multi-component salt solution spaces to achieve user-specified LCST targets. The platform demonstrates convergence to target properties within a minimal number of experiments. It strategically explores the parameter space, learns from informative "off-target" results, and self-corrects to achieve the final targets. By providing an accessible and adaptable blueprint, this work lowers the barrier to entry for autonomous experimentation and accelerates the design and discovery of functional polymers.}}\\%The abstract goes here instead of the text "The abstract should be..."

%%%END OF ABSTRACT%%%%

\rmfamily %Please do not remove this line.

%%%MAIN TEXT%%%%
\section*{Introduction}
Integrating artificial intelligence with laboratory automation has given rise to the self-driving laboratory (SDL) concept, a term first discussed by Isenhour in 1985\cite{isenhour1985robotics}. An SDL is a closed-loop system that enhances the scientific method by integrating automated hardware for executing experiments with intelligent software for data analysis and planning subsequent steps. This paradigm can accelerate scientific discovery, with demonstrated potential to reduce research timelines by factors of 10 to 1000\cite{delgado2023research, volk2024performance, hase2019next}, while also cutting costs, waste, and energy consumption by 10 to 100 times\cite{wang2023sustainable}. In addition to efficiency, SDLs enhance operational consistency, ensuring experimental repeatability and reducing human-induced variability\cite{tom2024self}. Foundational studies, such as King et al.'s Robot Scientist "Adam" \cite{king2009robot}, Granda et al.'s organic synthesis robot \cite{granda2018controlling}, Burger et al.'s mobile robotic chemist \cite{burger2020mobile}, and MacLeod et al.'s thin-film materials laboratory "Ada" \cite{macleod2020self}, have showcased SDL's capabilities in autonomously driving discoveries in chemistry and materials science. Recent comprehensive reviews highlight advancements, challenges, and efforts in SDL technologies across various applications, including battery research \cite{dave2021autonomous}, additive manufacturing \cite{snapp2021increasing}, nanoparticle synthesis \cite{volk2023alphaflow,abolhasani2023rise}, thin-film electronics \cite{macleod2020self}, and chemical reaction discovery \cite{granda2018controlling,burger2020mobile}. Looking ahead, the community is focused on addressing the key challenges inherent in establishing these complex systems \cite{seifrid2022autonomous}, with a growing emphasis on creating robust digital infrastructures and community standards to democratize access to SDL technology \cite{tom2024self,bayley2024autonomous,doloi2025democratizing}

Over the past five years, a rich open-source software ecosystem has emerged to democratize SDL operation and data handling.  Frameworks such as PyOpticon \cite{randall2025pyopticon} and IvoryOS \cite{zhang2025ivoryos} auto-generate graphical user interfaces around Python device classes, while AlabOS \cite{fei2024alabos}, HELAO-async \cite{guevarra2023orchestrating}, and ChemOS 2.0 \cite{roch2018chemos,sim2024chemos} provide scalable workflow management. These open-source tools empower researchers to construct and control automated workflows more easily, lowering the entry barrier to autonomous science.

\begin{figure*}[t]
  \centering
  \includegraphics[width=\textwidth]{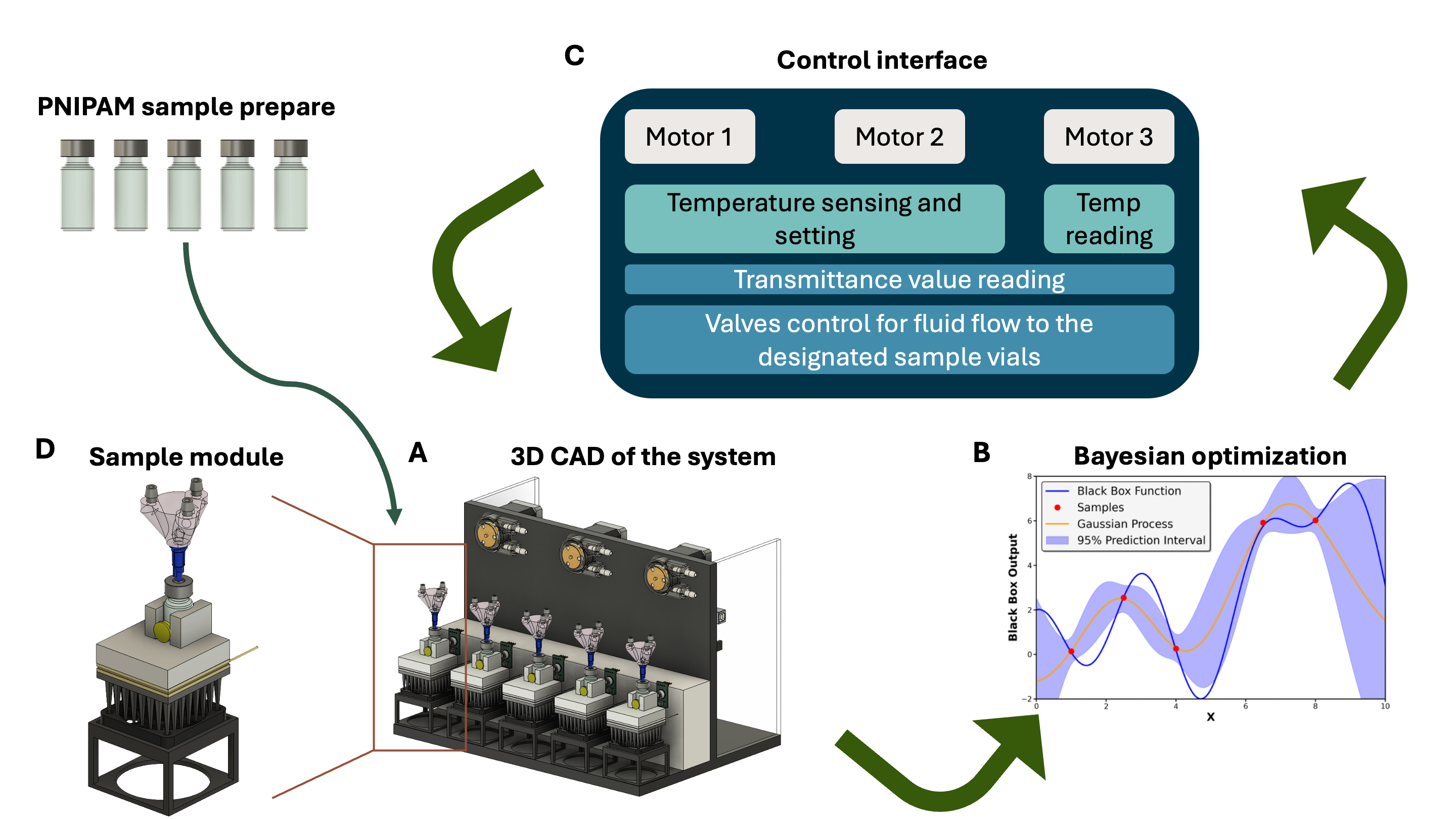}
  \caption{
    \textbf{Schematic of the Closed-Loop Autonomous Experimental Workflow.}
    (A) The view of the complete system, showing the parallel array of five modules and the integrated fluid-handling capability. (B) BO model update and recommend the most promising conditions for the subsequent experiment (C) Through the graphical user interface (GUI), a user specifies the volume of each stock solution to be dispensed, allowing for the precise formulation of each sample's final composition (D) A detailed view of a single module highlights the key components: a multi-port sample holder, photodiode detector, a Peltier element for rapid heating and cooling, and a heat sink for thermal stability. This assembly enables automated, in-situ measurement of LCST by controlling both sample composition and temperature.
  } 
  \label{Figure1}
\end{figure*}

Yet, the path to SDLs still faces challenges in hardware adaptability. Recent demonstrations show both the promise and the price of full hardware integration. At the high end, Hitosugi et al. built a dual-arm, dual-hexagon vacuum cluster that deposits, transfers, and characterizes Nb-doped $TiO_2$ films in a closed Bayesian optimization (BO) loop, yielding $\approx 12$ samples per day at $\approx2h$ per cycle\cite{shimizu2020autonomous}, but such chambers require a level of capital investment and infrastructure that is typically inaccessible to individual academic research groups. At the opposite end, the open-hardware community embraces low-cost, modular "frugal twins"—physical surrogates of high-end SDLs that capture core functionality at a fraction of the cost\cite{frugaltwins}. For example, the high cost of commercial high-throughput combinatorial printing systems, often exceeding \$100,000, puts them out of reach for most research groups. In contrast, the open-hardware "frugal-twin" offers a more accessible alternative. For example, the \textit{Archerfish} system demonstrates this principle by retrofitting a consumer 3-D printer for approximately \$500~\cite{siemenn2025archerfish}. Despite this 100-fold reduction in cost, the platform achieves a remarkable throughput of up to 250 unique material compositions per minute. Such dramatic cost compression is key to democratizing access to autonomous, data-rich experimentation and accelerating materials innovation. 

In this work, we develop a "frugal-twin" autonomous platform for soft-material discovery that closes the loop for the data-driven exploration of functional polymers. As a testbed for our autonomous platform, we chose poly(N-isopropylacrylamide) (PNIPAM), the most widely studied thermoresponsive polymer\cite{halperin2015poly}. In water, PNIPAM exhibits a sharp and reversible phase transition at its lower critical solution temperature (LCST), typically around 32 \textdegree C. Below this temperature, the polymer chains exist as soluble, extended coils. Above it, they collapse into compact, hydrophobic globules. The proximity of this transition to physiological temperatures makes PNIPAM a candidate material for developing "smart" technologies, including controlled drug delivery systems\cite{mitchell2015poly,ayar2021rechargeable}, tissue engineering scaffolds\cite{ashraf2016snapshot, wu2023investigation}, and biosensors\cite{das2024poly,islam2014poly}. 

A critical feature of PNIPAM is that its LCST is tunable rather than fixed. It can be tuned by varying the concentration and type of dissolved salts\cite{du2010effects}. This phenomenon is governed by the Hofmeister series, where different ions modulate the phase transition by altering the structure of the water surrounding the polymer chains \cite{burba2008salt}. Salts can induce a "salting-out" effect, disrupting the polymer's hydration shell and promoting hydrophobic interactions, which typically lowers the LCST. Mechanistically, anions tend to exert a more pronounced influence on the LCST than cations, in line with their established Hofmeister rankings\cite{pastoor2015cation}, which orders ions by their "salting-out" strength; for instance, strongly salting-out ions like sulfate cause a much greater LCST depression than weakly interacting ions like chloride.

The behavior in multi-component solutions containing mixtures of two or three salts is more complex. These mixtures can exhibit synergistic effects that are difficult to predict, as the cations and anions compete to stabilize or destabilize the polymer's hydration shell. This complexity presents both a challenge and an opportunity: using salt mixtures offers more degrees of freedom for fine-tuning the LCST, but the vast, non-linear design space makes optimization more difficult.

In our "frugal-twin" autonomous platform, we demonstrate its capability by deploying BO to navigate the multi-component salt solution space of PNIPAM and precisely control its LCST by tunning the types and concentrations of the salts. The results indicate that our integrated robotic and machine learning system can achieve desired material properties with high efficiency and minimal human intervention, moving toward the accelerated and autonomous discovery of functional polymers with cost-effective systems.
\section*{Experimental}
\subsection*{Hardware}
The automated platform consists of five identical modules, integrated with a centralized fluid-handling system for automated sample preparation and analysis, as shown in Figure 1a. This modular architecture enables simultaneous, independent experiments. Each module works as an independent reactor and is designed to allow the automated, in-situ measurement of LCST. To determine the LCST, each module has an optical detection system consisting of a 600 nm LED and a photodiode detector (FDS100) positioned across the sample vial. During a measurement cycle, the temperature is ramped from a starting point below the expected LCST to a final point above it. The detector records the light intensity passing through the sample at each temperature step. The LCST is then defined as the temperature at which the normalized transmittance reaches 50\% of its initial value, determined by linear interpolation between the two data points that span this threshold. A detailed view is shown in Figure 1D, highlighting the key components: a multi-port sample holder, photodiode detector, a Peltier TEC-12706AJ thermoelectric module for heating and cooling, and a heat sink to ensure thermal stability. The system's Arduino Mega 2560 microcontroller implements a Proportional-integral-derivative (PID) algorithm to precisely regulate the sample temperature, using feedback from a non-contact infrared sensor to modulate power to the thermoelectric module. The fluid-handling subsystem is responsible for creating samples with specific chemical compositions. This is achieved using a set of DC motor-driven peristaltic pumps, managed by DRV8825 stepper motor drivers, for accurate liquid dispensing from source containers. Solenoid valves, actuated via a switching circuit using TN0106N3 MOSFETs, control the direction of fluid flow to the designated sample vials. The Arduino serves as the central controller for the entire platform, synchronizing the operation of the pumps, valves, and thermoelectric modules, and providing integrated control over both liquid proportioning and temperature control based on user-defined parameters.

\subsection*{Software}
The automated liquid distribution system is controlled by a Python application with a graphical user interface (GUI) developed with the PyQt6 framework, which allows users to monitor and manage different components and processes in the system directly. Through the GUI, users can manage liquid dispensing, actuate valves for fluid distribution, and set and monitor temperature parameters for each sample. The software includes a data analysis module that allows users to select experimental data folders, visualize temperature profiles, and calculate the LCST. 

The program combines custom scripts and modules, leveraging the Python Standard Library alongside several third-party open-source libraries. Core functionalities include hardware interfacing via \verb|pyfirmata| for Arduino communication. The software architecture comprises separate functionalities, including motor control, temperature regulation, valve operation, and data visualization. This modular design makes it easier to organize and expand the codebase. The Python code for system control and data analysis is available on GitHub. 

\subsection*{Chemicals and materials}

The precursor monomers and initiator solutions were first prepared to synthesize PNIPAM hydrogels. N-isopropylacrylamide (NIPAM,97\%, Sigma-Aldrich) was dissolved in deionized water (18.2~M$\Omega\cdot$cm)  at 0.806 g per 10 ml. The crosslinker, N,N'-methylenebisacrylamide (BIS, 97\%, purchased from Fisher Scientific), was prepared as a solution of 0.2196 g in 10 mL of deionized water. The initiator, potassium persulfate (KPS, 99. 0\%, Sigma-Aldrich), was dissolved in deionized water at 0.135 g per 10 ml concentration. N,N,N',N'-Tetramethylethylenediamine (TEMED, >98.0\% (GC), TCI America, purchased from Fisher Scientific) was used as an accelerator and added directly as received. For a typical hydrogel synthesis, the following volumes of the prepared stock solutions and reagents were combined: 643 µL of the NIPAM solution, 85 µL of the BIS solution, 121 µL of the KPS solution, and 8.9 µL of TEMED. 500 µL of deionized water was added to reach the final volume. After polymerization, a gentle  thermal treatment was used to separate the hydrogel from the surrounding liquid, which was then removed.

\section*{Results and discussion}
\begin{figure*}[htbp]
  \centering
  \includegraphics[width=\textwidth]{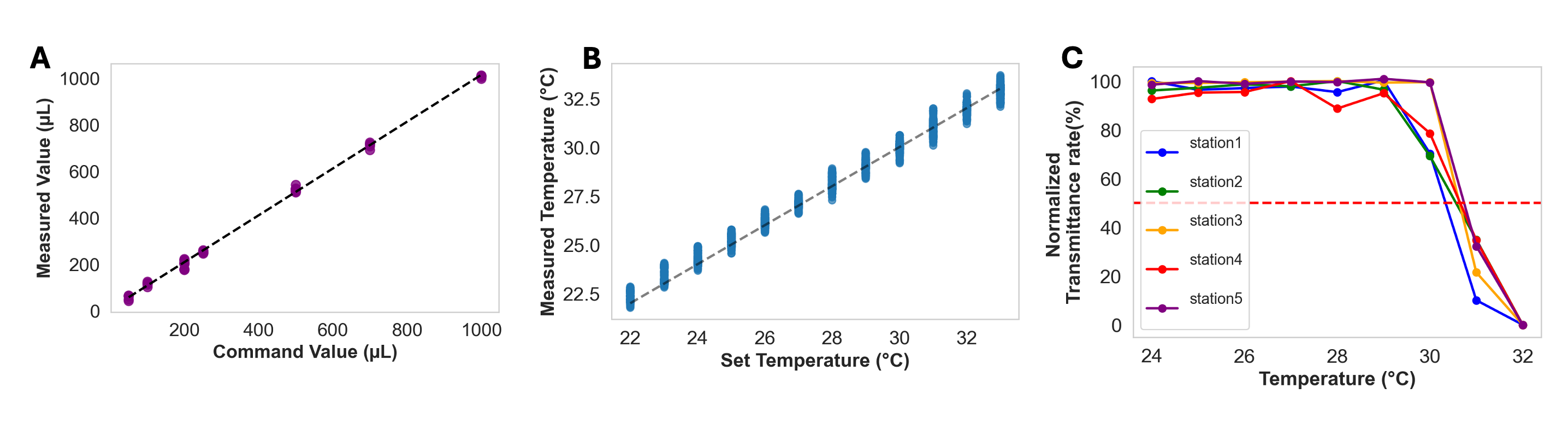}
  \caption{
    \textbf{System validation and performance testing.}
    (A) Liquid handling accuracy and precision, showing measured versus commanded volumes for automated dispensing. 
    (B) Temperature control stability, showing measured temperatures versus set temperatures across the experimental range. Measurements were recorded every 3 seconds for 8 minutes at each setpoint, with an average standard deviation of 0.30\textdegree C, confirming robust and precise temperature control.
    (C) Repeatability of standard PNIPAM LCST measurements performed at five independent sample modules, with an average LCST of 30.59\textdegree C and a standard deviation of 0.15\textdegree C. 
    These results demonstrate the automated platform's accuracy, temperature stability, and measurement reproducibility.
  } 
  \label{Figure2}
\end{figure*}
\subsection*{Validation of Fluid Handling, Temperature Control, and Measurement Repeatability}
To establish a reliable and reproducible experimental platform, the system's performance was characterized in three key areas: liquid handling precision, temperature control stability, and LCST measurement repeatability. First, the precision of the fluid handling system was confirmed over a broad range (20–1000 µL). As shown in Figure 2A, the measured values closely matched the commanded setpoints, yielding high linearity, $R^2=0.998$, and demonstrating accurate fluid delivery. The platform's temperature stability was assessed across a range of setpoints relevant to PNIPAM LCST transitions. For each setpoint, the temperature was measured using an IR sensor every 3 seconds for 8 minutes. Figure 2B shows the platform's thermal conditions and achieves accurate temperature control throughout the entire experimental range, with an average standard deviation across all setpoints of 0.30\textdegree C, demonstrating robust and precise thermal regulation essential for reproducible LCST measurements. Finally, the overall LCST measurement repeatability was evaluated by performing standard PNIPAM LCST determination at five independent sample modules. As illustrated in Figure 2C, the measured LCSTs across stations exhibited good consistency, with an average value of 30.59\textdegree C and a standard deviation of 0.15\textdegree C. Collectively, these results show that the automated platform delivers reproducible results, establishing a reliable foundation for subsequent autonomous experimentation and machine learning-guided optimization.

\subsection*{Autonomous experimentation with machine learning model}
\begin{figure*}[t]
  {\centering
  \includegraphics[width=0.65\textwidth]{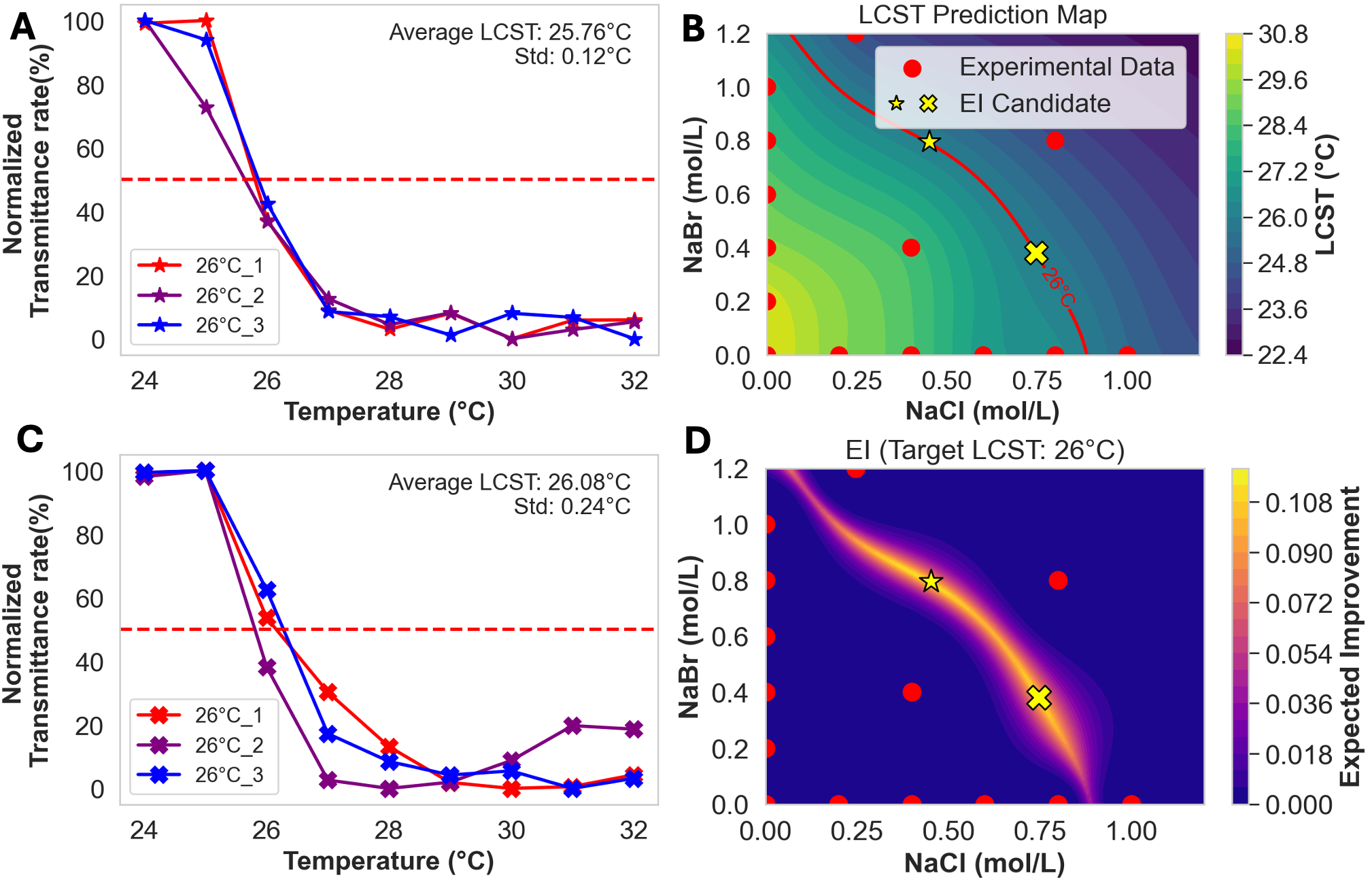}
  \caption{
  Machine Learning-guided optimization of LCST in two-salt compositions.
    (\textbf{A}) The result from the second round of BO (star) yields an LCST of $25.76 \pm 0.12$\textdegree C, closely approaching the target. (\textbf{B, D}) After this measurement, the GPR model is updated, generating a new LCST prediction map (\textbf{B}) and the corresponding EI landscape (\textbf{D}) to guide the third round. The EI map directs the platform to select the next candidate (cross) from the region of highest predicted value. (\textbf{C}) The third round experiment at this new BO-recommended location achieves the target with a measured LCST of $26.08 \pm 0.24$\textdegree C.}
  }
  \label{Figure3}
\end{figure*}
To optimize the LCST of PNIPAM in two- and three-salt solutions, we implemented a machine learning-driven experimental loop guided by BO\cite{wang2022bayesian}. This data-driven approach has proven highly effective and data-efficient across a range of materials discovery challenges, such as for thermoelectric materials\cite{song2024machine, saeidi2022machine}, additive manufacturing\cite{wang2023physics}, and jet sintering\cite{cheng2024bayesian}. 

For our surrogate model, we chose Gaussian Process Regression (GPR) due to its ability to provide not only predictions but also the quantification of uncertainty. The GP defines a probability distribution over functions, and our model assumes the $LCST (f(\boldsymbol{x})))$ can be represented as:
\begin{equation}
    f(\boldsymbol{x}) \sim \mathcal{GP}(m(\boldsymbol{x}), k(\boldsymbol{x}, \boldsymbol{x'}))
\end{equation}
where $m(\boldsymbol{x})$ is the mean function (assumed to be zero) and $k(\boldsymbol{x},\boldsymbol{x'})$ is the covariance function, or kernel, which models the relationship between different points in the parameter space.

The experimental framework comprises two models that operate on input vectors, $\boldsymbol{x}$, representing either two-dimensional ($\boldsymbol{x} = [C_{\text{NaCl}}, C_{\text{NaBr}}]$) or three-dimensional salt concentrations ($\boldsymbol{x} = [C_{\text{NaCl}}, C_{\text{NaBr}}, C_{\text{CaCl}_{2}}]$). The LCST output, $y$, for each composition is the mean of three replicate experiments. To improve numerical stability, each feature and the response are standardized to zero mean and unit variance:
\begin{equation}
 \tilde{\mathbf{x}} = \frac{\mathbf{x}-\boldsymbol{\mu}_{\mathbf{x}}}{\boldsymbol{\sigma}_{\mathbf{x}}}, \qquad \tilde{y} = \frac{y-\mu_{y}}{\sigma_{y}},
 \label{eq:standardization}
\end{equation}

where $\mu_x$ and $\sigma_x$ are the mean vector and standard deviation vector. 
% Make sure you have \usepackage{amsmath} in your preamble for \substack to work!
\begin{equation}
    \begin{aligned}
    k_{\text{2-salt}}(\boldsymbol{\tilde{x}},\boldsymbol{\tilde{x}}') = & \underbrace{\sigma_{f}^{2}\,\text{Matern}_{\nu=1}\bigl(\lVert\boldsymbol{\tilde{x}}-\boldsymbol{\tilde{x}}'\rVert;\,\boldsymbol{\ell}\bigr)}_{\substack{\sigma_{f}^{2} = 12.25 \\ \boldsymbol{\ell} = [12.3, 15.3]}} \\
    & + \underbrace{\sigma_n^{2}\,\delta_{\boldsymbol{\tilde{x}},\boldsymbol{\tilde{x}}'}}_{\sigma_n^{2} = 6.7 \times 10^{-12}}
    \end{aligned}
    \label{eq:k_2salt_annotated}
\end{equation}
For the two-salt surrogate model (NaCl, NaBr), we adopted a composite kernel combining Matern and White Noise kernel. The model's hyperparameters were optimized by maximizing the log-marginal likelihood, and the final values reveal key features of the system's behavior. The fitted signal variance is \(\sigma_f^{2}=12.25\), which sets the overall amplitude of LCST fluctuations captured by the smooth Matern component. The anisotropic length-scales are \(\ell_{\text{NaCl}}=12.3\) and \(\ell_{\text{NaBr}}=15.3\) (in standardized units), which suggests that the LCST is more sensitive to changes in NaCl concentration compared to the more gradual influence of NaBr. A smoothness parameter of $\nu=1$ was chosen for the Matern kernel to balance model complexity with the limited dataset. The performance of the GPR model was evaluated using leave-one-out cross-validation. The resulting parity plot, presented in Supplementary Information (SI) Figure S1, shows a strong correlation between the predicted and experimental values. The model exhibits high predictive accuracy, achieving a coefficient of determination ($R^{2}$) of 0.941.

For the higher-dimensional three–salt space (NaCl, NaBr, CaCl$_2$), a more complicate composite kernel was required to accurately model the LCST response. 
\begin{equation}
    \begin{aligned}
    k_{\text{3-salt}}(\boldsymbol{\tilde{x}},\boldsymbol{\tilde{x}}') = & \underbrace{\sigma_{\text{lin}}^{2}\,(\boldsymbol{\tilde{x}}^{\top}\boldsymbol{\tilde{x}}')}_{\approx 1.35} \\
    & + \underbrace{\sigma_{\text{RBF}}^{2}\,\exp\Biggl(-\frac{1}{2}\sum_{d=1}^{3}\frac{(\tilde{x}_d-\tilde{x}'_d)^2}{\ell_{\text{RBF},d}^2}\Biggr)}_{\substack{\sigma_{\text{RBF}}^{2} \approx 0.107 \\ \boldsymbol{\ell}_{\text{RBF}} = [0.334, 1.87, 0.535]}} \\
    & + \underbrace{\sigma_{\text{Mat}}^{2}\,\text{Matern}_{\nu=2.5}\bigl(\lVert\boldsymbol{\tilde{x}}-\boldsymbol{\tilde{x}}'\rVert;\,\boldsymbol{\ell}_{\text{Mat}}\bigr)}_{\substack{\sigma_{\text{Mat}}^{2} \approx 0.061 \\ \boldsymbol{\ell}_{\text{Mat}}=[0.0139, 100, 100]}} \\
    & + \underbrace{\sigma_n^{2}\,\delta_{\boldsymbol{\tilde{x}},\boldsymbol{\tilde{x}}'}}_{\sigma_n^{2} = 10^{-10}}
    \end{aligned}
    \label{eq:k_3salt_final_all_values}
\end{equation}

This complex kernel provides a robust model by simultaneously capturing a dominant linear trend via a Dot Product component and modeling complex, non-linear interactions with RBF and Matern terms. 

\begin{figure*}[htbp]
    \centering
    \includegraphics[width=\textwidth]{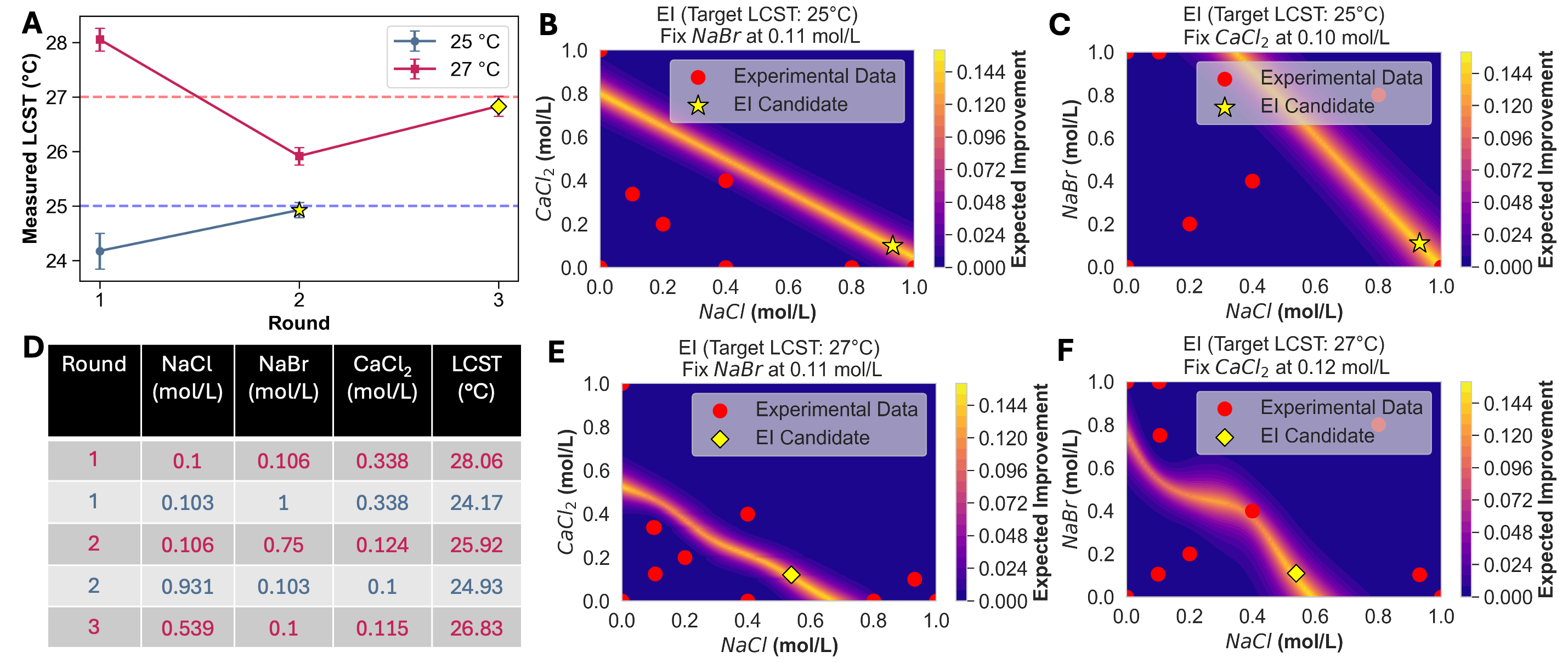}
    \caption{\textbf{Machine learning-guided optimization of LCST in three-salt compositions.} (\textbf{A}) Optimization trajectories for targeting 25\textdegree C (blue) and 27\textdegree C (red). (\textbf{B, C, E, F}) Representative EI heat maps visualizing the model's decision-making process. The algorithm identifies optimal candidates (yellow markers) by targeting regions predicted to offer the greatest improvement over the current best observation. Red circles denote previously measured experimental data points. (\textbf{D}) Tabulated salt concentrations (NaCl, NaBr, CaCl$_2$) and resulting LCSTs for each experimental round.}
    \label{Figure4}
\end{figure*}

These trained surrogate models are integrated into a BO framework, utilizing an expected improvement (EI) acquisition function. As the objective is to identify a specific target LCST rather than a global optimum, the standard EI function is adapted. For this target-seeking problem, 'improvement', $I(x)$, is mathematically defined as the negative absolute difference between the model's predicted LCST and the specified target value. 
\begin{equation}
I(x) = -|y_{\text{pred}}(x) - y_{\text{target}}|
\end{equation}
This formulation directs the optimization process toward regions of the parameter space predicted to be closer to the target.
\begin{equation}
  \text{EI}(x) = I(x)\Phi(z) + \sigma(x)\phi(z)
\end{equation}

\noindent where:
\begin{equation}
  z = \frac{-|y_{\text{pred}}(x)-y_{\text{target}}|}{\sigma(x)}
\end{equation}
and $\Phi(\cdot)$ and $\phi(\cdot)$ are the standard normal CDF and PDF, respectively.
EI not only accounts for the predicted improvement but also considers the uncertainty of the prediction, encouraging the exploration of less certain areas that still have potential. To determine the optimal experimental conditions, we evaluate EI across a dense grid of candidate concentrations and select the condition that maximizes EI for subsequent experimental validation. After each optimization round, the corresponding GPR model is retrained using the expanded dataset, and subsequent experimental conditions are selected based on maximum EI values. By iteratively updating the GP model and recalculating EI after each new experiment, our autonomous platform effectively navigates the chemical parameter space toward rapid convergence to desired LCST values with minimal experimental overhead.

We first deployed our autonomous platform on the two-salt system (NaCl–NaBr) to perform a targeted optimization for an LCST of 26\textdegree C. Before initiating the closed-loop process detailed in Figure 3, the GPR model was initialized with a 13-point experimental dataset to provide broad coverage of the two-salt parameter space. This initial set was selected using a structured approach, sampling points along each axis and an interior point to capture a combined effect. Initially, the model-guided experiments yielded an average LCST of $24.55 \pm 0.32$\textdegree C, below the intended target (Figure S2). Figure 3A shows normalized transmittance data from the BO-recommended salt condition (NaCl: 0.452 mol/L, NaBr: 0.796 mol/L), resulting in LCST measurement of $25.76 \pm 0.12$\textdegree C, approaching the target. The updated GPR model provided a new prediction of the LCST landscape (Figure 3B). Based on this updated model, the EI acquisition function is used to identify the optimal candidate for the next round of experiments. Figure 3C illustrates the subsequent experimental validation at the next BO-selected concentration (NaCl: 0.748 mol/L, NaBr: 0.38 mol/L), achieving an LCST of $26.08 \pm 0.24$\textdegree C, within the acceptable margin of the target. Figure 3D visualizes the corresponding EI landscape for the 26\textdegree C LCST target, clearly highlighting high-potential regions (bright yellow) used by the algorithm to efficiently steer experimentation toward optimal conditions. Together, these iterative results demonstrate how the autonomous experimental platform effectively leverages BO to guide experiments, achieving rapid convergence to the targeted LCST with minimal experimental effort. To confirm the general applicability and robustness of our system, additional validation experiments targeting LCSTs of 27\textdegree C and 30\textdegree C were conducted, successfully identifying appropriate salt compositions. Detailed results of these validations are provided in Figure S3.

\begin{figure*}[t]
    \centering
    \includegraphics[width=\textwidth]{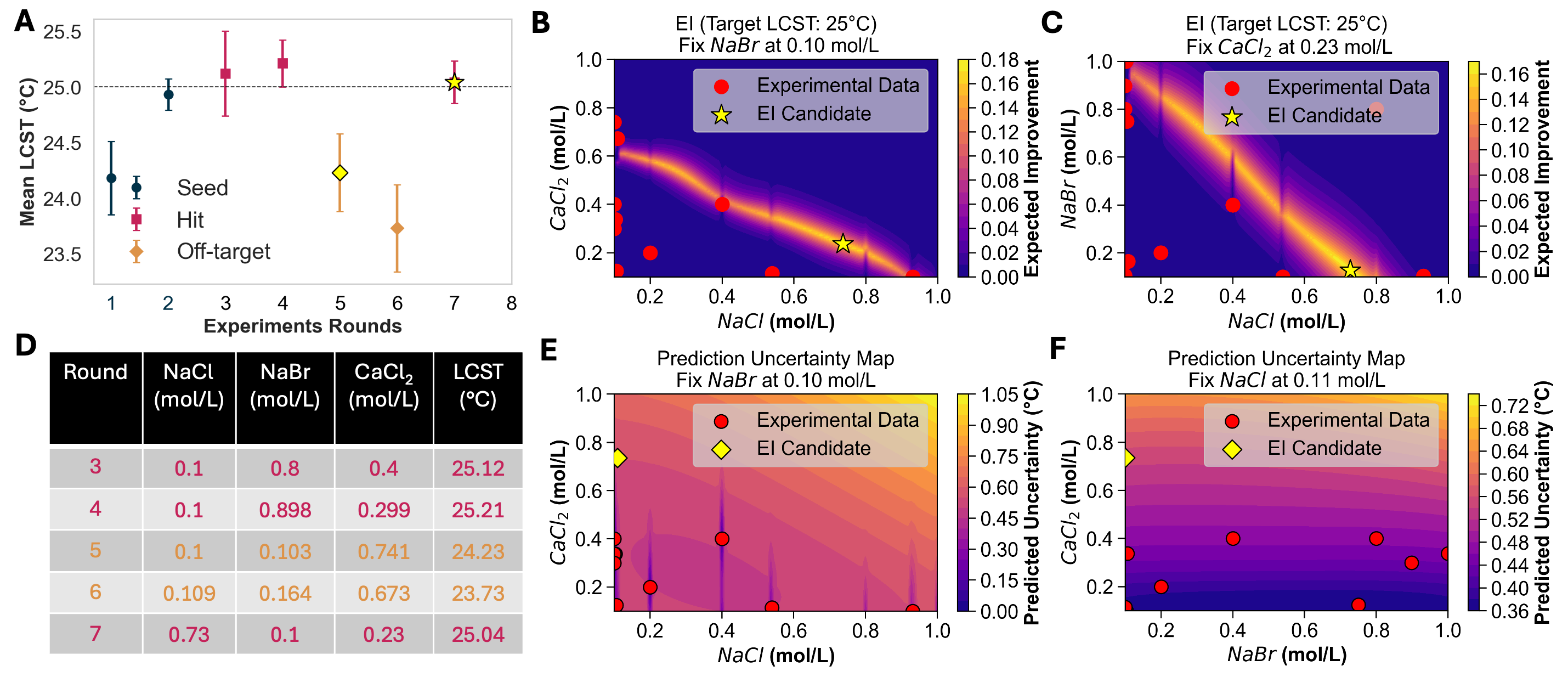}
    \caption{\textbf{Five-round verification experiment targeting on 25\textdegree C LCST} (\textbf{A}) The optimization trajectory for a five-round verification campaign (rounds 3-7). After two successful initial hits (purple square), the system makes two exploratory off-target measurements (orange diamond) and then successfully self-corrects to achieve a precise final hit on the 25\textdegree C target in round 7 (yellow star). (\textbf{B, C}) EI maps generated for round 7, with the star highlighting the selected EI candidate. (\textbf{E, F}) The corresponding prediction uncertainty maps illustrate how the model's confidence changes throughout the experiment. The yellow diamond marks the round 5 candidate, which was specifically chosen on regions of high uncertainty. (\textbf{D}) Tabulated salt compositions and resulting LCSTs for each verification round.}
    \label{Figure5}
\end{figure*}

To demonstrate the efficacy of our closed-loop optimization framework in a more complex parameter space, we used the three-salt GPR model (NaCl, NaBr, and CaCl$_2$), initialized with seven diverse data points, which was strategically selected to provide a robust starting point for the optimization. This initial design of experiments included points along each axis to capture the effect of each individual salt, as well as interior points representing both two- and three-salt mixtures. This structured approach, chosen over random sampling, ensures the initial model is informed by a diverse set of conditions. The platform was used to perform a series of experiments aimed at identifying salt concentrations that met two distinct LCST targets, as illustrated in Figure 4. Task 1, targeting an LCST of 25\textdegree C, demonstrated the platform's power for rapid, iterative refinement. The first BO-guided experiment (round 1) yielded an LCST of $24.17\pm 0.33$\textdegree C (Fig. 4D). While not at the target, this first step provided the model with information about the local parameter landscape. As illustrated in the SI, Figure S4, this new information significantly refined the GPR model, correcting its initially simplified prediction of the 25°C contour. With this updated understanding of the system, the platform's EI acquisition function was then able to propose a more refined composition for the next iteration. This second BO-guided experiment (round 2) successfully achieved an LCST of $24.93\pm 0.14$\textdegree C , which within the error margin of the target. This process demonstrated the platform's capacity to learn from each experiment and direct it towards the desired goal (Fig. 4A, blue trace). 

Task 2, targeting a 27\textdegree C LCST, highlighted the platform's ability of exploration for complex problems. The initial experiment (round 1) yielded an LCST of $28.06\pm0.21$\textdegree C. The BO-guided experiment conducted in round 2 yielded a temperature of $25.92\pm0.16$\textdegree C. While neither experiment hit the target, these results provided valuable data from a different region of the parameter space. After incorporating the learnings from the previous two rounds, the GPR model generated the updated Expected Improvement landscape for the third round, as shown in Figures 4E and 4F. Guided by this heat map, the platform selected its candidate for round 3, which achieved an LCST of $26.83\pm0.18$\textdegree C. This brought the measurement to within 0.2\textdegree C of the target. The completion of these two tasks shows the robustness and efficiency of the autonomous platform. 

To further test the platform's capabilities, we conducted a five-round verification experiments targeting at 25\textdegree C LCST. Those experiments (rounds 3-7) were designed to show the system's ability to self-correct and navigate a complex local landscape, as illustrated in Figure 5. The two successful experiments in round 3 and round 4 (Fig. 5A, purple squares), which confirmed the existence of a viable compositional region near the 25°C target. Following this initial success, the platform strategically shifted from exploitation to exploration. The prediction uncertainty maps revealed regions where the model's confidence was low, and guided by this, the algorithm intentionally selected a candidate in a high uncertainty area for round 5 (orange diamond, Fig. 5E, F). This exploratory approach, along with an additional investigation in round 6, resulted in LCST values of $24.23\pm0.35$ \textdegree C and $23.73\pm0.39$ \textdegree C, both were off the target.

From those failures, the exploratory data points were crucial for building a more accurate GPR model for the uncertain regimes in the parameter space. After incorporating the points of both successful experiments and informative misses, the platform recommends the final EI landscape shown for round 7 (Fig. 5B, C). With a much richer understanding of the system, the algorithm made its final selection (yellow star). This last experiment achieved an LCST of $25.04\pm0.19$\textdegree C, which demonstrates the system's ability to self-correct after exploratory steps and converge on the target. These experiments demonstrate the platform's ability to navigate the exploration-exploitation trade-off, which is critical for efficiently mapping a complex parameter space and rapidly converging on desired experimental results.

\section*{Conclusions}
In this work, we have successfully designed, built, and demonstrated an autonomous closed-loop platform for the accelerated optimization of poly(N-isopropylacrylamide)'s LCST. We demonstrated that by integrating robotic fluid handling with a BO framework utilizing a GPR surrogate model, our system can efficiently navigate complex, multi-component chemical spaces.

The comprehensive validation of the platform's hardware confirmed its high precision in liquid handling, stability in temperature control, and reproducibility, establishing a reliable framework for data-driven discovery. The machine learning-guided workflow was shown to be effective for optimizing the LCST of PNIPAM in complex, multi-component salt solutions. In the experiments, the system converged on specified LCST targets in both two- and three-salt systems, showcasing its adaptability to problems of varying complexity.

We also demonstrated the platform's ability to self-correct. The multi-round verification experiments highlighted how the system strategically balances exploitation of known information with the exploration of uncertain regions. By intelligently learning from both on-target experiments and informative off-target experiments, the platform successfully recovered from deviations to achieve a final, high-precision objective. This capacity to leverage all experimental outcomes for model improvement is fundamental to the SDL's stability, enabling it to converge on targets even after exploring non-optimal conditions.

% \section*{Author Contributions}
% We strongly encourage authors to include author contributions and recommend using \href{https://casrai.org/credit/}{CRediT} for standardised contribution descriptions. Please refer to our general \href{https://www.rsc.org/journals-books-databases/journal-authors-reviewers/author-responsibilities/}{author guidelines} for more information about authorship.

\section*{Conflicts of interest}
There are no conflicts to declare.

\section*{Data availability}

All experimental code supporting the findings of this study is openly available in the associated GitHub repository: \href{https://github.com/An-xu/SDL_LCST_OP}{https://github.com/An-xu/SDL\_LCST\_OP}.
The repository contains the source code for automated control and data analysis.

%%%END OF MAIN TEXT%%%

%  For footnotes in the main text of the article please number the footnotes to avoid duplicate symbols. e.g.  \footnote[num]{your text} the corresponding author \ast counts as footnote 1, ESI as footnote 2, e.g. if there is no ESI, please start at [num]=[2], if ESI is cited in the title please start at [num]=[3] etc. Please also cite the ESI within the main body of the text using \dag.

% The \balance command can be used to balance the columns on the final page if desired. It should be placed anywhere within the first column of the last page.

% \balance

% If notes are included in your references you can change the title from 'References' to 'Notes and references' using the following command:
% \renewcommand\refname{Notes and references}

%%%REFERENCES%%%

\bibliographystyle{unsrtnat}   % or plainnat, ieeetranN, etc.
\bibliography{rsc}

\end{document}